\documentstyle[preprint,aps]{revtex}

\def\picture #1 by #2 (#3){
  \vbox to #2{
    \hrule width #1 height 0pt depth 0pt
    \vfill
    \special{picture #3} 
    }
  }

\def\scaledpicture #1 by #2 (#3 scaled #4){{
  \dimen0=#1 \dimen1=#2
  \divide\dimen0 by 1000 \multiply\dimen0 by #4
  \divide\dimen1 by 1000 \multiply\dimen1 by #4
  \picture \dimen0 by \dimen1 (#3 scaled #4)}
  }

\begin{document}

\title{How important is the three-nucleon force?}

\author{T-Y. Saito and I. R. Afnan}
\address{School of Physical Sciences,
        The Flinders University of South Australia,\\
        Bedford Park, SA 5042, Australia}
\date{\today}

\maketitle

\begin{abstract}

By calculating the contribution of the $\pi-\pi$ three-body force to the
three-nucleon binding energy in terms of the $\pi N$ amplitude  using
perturbation theory, we are able to determine the contribution of the
different $\pi N$ partial waves to the three-nucleon force. The division of
the $\pi N$ amplitude into a pole and nonpole gives a unique procedure for
the determination of the $\pi NN$ form factor in the model. The total
contribution of the three-body force to the binding energy of the triton is
found to be very small.

\end{abstract}


\newpage

The discrepancy between the results of the exact calculation of the binding
energy of the triton using a number of realistic nucleon-nucleon potentials,
and the experimental value of 8.48~MeV, has been an outstanding problem in
nuclear physics for a number of years. A commonly accepted solution has been
the introduction of a three-nucleon force that will bridge the gap between
the calculated binding energy \cite{IS86,FG88}, based on two-body
interaction, and the experimental binding energy. The origin of such a
three-body force lies in the fact that the nucleons are treated as point
particles interacting via a two-body potential that is often assumed to be
local. This approximation neglects the contribution of the mechanisms whereby
one of the nucleons emits a meson that scatters off a second nucleon and then
gets absorbed onto the third nucleon; see Fig.~\ref{Fig.1}. Thus in a theory
where the meson degrees of freedom are suppressed, one needs to include the
contribution from the three-body force given in Fig.~\ref{Fig.1}.

Over the past ten years there have developed two main approaches to the
determination of this three-nucleon interaction from an underlying
meson-nucleon theory. One of the first such three-body forces was developed by
the Tucson-Melbourne (TM) group\cite{CS79,CG81,CP93}. They determined the
$\pi N$ amplitude that goes into the calculation of the three-body force by
emphasizing the role of symmetry~\cite{BG68}, and in particular, stressed the
importance of using current algebra and soft pion theorems, to fix the
behavior of the $\pi N$ t-matrix, off-mass shell and near the $\pi N$
threshold. This amplitude was then used to calculate the three-body force
given in Fig.~\ref{Fig.1}.  The second
approach~\cite{HS83a,HS83b,Sa86,PH90,SS92,PS93,PR91} emphasized the fact that
the $\pi N$  amplitude is dominated at medium energies by the
$\Delta(1230)$ resonance, which is considered as the first excited state of
the nucleon. This suggests that one should introduce coupling between the
$NN$, $N\Delta$, and $\Delta\Delta$ channels, and solve the coupled channel
problem for the $BBB$ system, where the baryon $B=N,\Delta$. In this way the
$\Delta$ component of the three-nucleon force is automatically included
in the three baryon system.

To motivate the need to reexamine the above approaches to the  three-nucleon
force, we should recall that the $NN$ amplitude, needed to calculate the
binding energy of the three-nucleon system using the Faddeev equations, is
$t^{NN}_\alpha(k,k';E_{NN})$, where $\alpha$ labels the partial wave, and the
energy of the $NN$ system, $E_{NN}$, is restricted to the domain
$-\infty<E_{NN}<-E_T$, with $E_T$ being the binding energy of the
three-nucleon system. The fact that we need this amplitude over the above
specified energy domain is a result of the fact that in the three-nucleon
system, the total energy is fixed at $E=-E_T$, and the spectator particle can
have any energy from zero to $\infty$. If we now examine, in the same spirit,
the contribution of the $\pi N$ amplitude to the three-body force, then the
{\it off-energy-shell} partial wave amplitudes we require are $t^{\pi
N}_\alpha(k,k';E_{\pi N})$ for all energies in the domain $-\infty<E_{\pi
N}<(m_N-E_T)$, where in this case we have included the rest mass of the
nucleon and pion in the $\pi N$ energy $E_{\pi N}$.
Here, the need for the $\pi N$ amplitude in this energy domain is a result of
the fact that the total energy is still $-E_T$, and the two spectator
nucleons have their full range of kinetic energy.
As a result, we have an integral of the $\pi N$ amplitude over this energy
domain when we calculate the contribution of the three-nucleon force to the
binding energy of the triton.
The second question that needs to be examined is that of the relative
contribution of the different $\pi N$ partial waves to the binding energy. In
the $NN-N\Delta$ coupled channel approach, it is assumed that the $P_{33}$ is
the only partial wave contributing to the three-body force, while in the TM
approach it is not clear how much of the $P$-wave contribution is included,
since the $\pi N$ amplitude is fixed by the low energy experimental data and
soft pion theorems.  Finally, we need to consider the $\pi NN$ form factor
used in Fig.~\ref{Fig.1} for the initial emission and final absorption of the
pion.  Traditionally, the $\pi NN$ form factor is taken to be a function of
the pion momentum only. This is a result of assuming that the nucleons are
{\it on-mass-shell} while the pion is {\it off-mass-shell}. At the $NN$
level, the cutoff mass in the $\pi NN$ form factor is determined by fitting
the $NN$ data, while for the three-nucleon force this cutoff mass is treated
either as a parameter~\cite{IS86,FG88}, or adjusted~\cite{CP93} to fit the
Goldberger-Treiman relation~\cite{CS90}.

For the present analysis we have taken the approach used in the formulation
of the $NN-\pi NN$ problem\cite{AB80,AM83} in which the off-energy-shell
$\pi N$ amplitude is divided into a pole and nonpole contribution. The
nonpole contribution gives the off-energy-shell $\pi N$ amplitude that
describes the $\pi N$ scattering in Fig.~\ref{Fig.1}. The removal of the
nucleon pole contribution from the $\pi N$ amplitudes avoids the double
counting of the one-pion exchange. The residue of the off-energy-shell
$\pi N$ amplitude at the nucleon pole defines the pion absorption and
production vertex in Fig.~\ref{Fig.1}. The corresponding dressed $\pi NN$
form factor is a function of the total $\pi N$ energy and the relative $\pi
N$ momentum. To maintain consistency with the three-nucleon wave function
(i.e, point nucleons), and avoid the problem of undercounting\cite{SSF85} due
to incomplete nucleon dressing\cite{B92}, we have used the $\pi NN$ form
factor extracted from the $\pi N$ amplitude rather than that extracted from
the $NNN-\pi NNN$ equations\cite{AM83b,CCS93}.

To simplify our model for the $\pi N$ amplitude, which is a solution of the
Lippmann-Schwinger equation, we can choose the $\pi N$ potential in the
$P_{11}$ channel to be the sum of an $s$-channel pole diagram and a background
term that is represented by a separable potential~\cite{MF81,MA82,MA85}. For
all other $S$- and $P$-wave channels we have used the separable potential of
Thomas\cite{Th76}. This potential, constructed for use in $\pi-d$ elastic
scattering, was adjusted to fit the $\pi N$ phase shifts and scattering
lengths in all $S$ and $P$ waves except the $P_{11}$. This potential gives a
partial wave amplitude that is of the form
\begin{equation}
t_\alpha(k,k';E) = g_\alpha(k)\,\tau_\alpha(E)\,g_\alpha(k')\ .  \label{eq:1}
\end{equation}
For the $P_{11}$ partial wave, where the $s$-channel nucleon pole plays an
important role, we have used the $\pi N$ potentials of McLeod and
Afnan\cite{MA85}. These potentials were developed for, and used in $\pi-d$
scattering and pion absorption on the deuteron~\cite{AM85}. These potentials
give a $\pi N$ scattering amplitude that is a sum of a pole term and a
nonpole term, i.e.,
\begin{eqnarray}
t_\alpha(k,k';E) &=& f^R(k,E)\,d^R(E)\,f^R(k',E) \nonumber \\
& & + g_\alpha(k)\,\tau_\alpha(E)\,g_\alpha(k')\quad\mbox{for}\
\alpha=P_{11}\                                                  \label{eq:2}
\end{eqnarray}
where the second term in Eq.~(\ref{eq:2}) is the nonpole amplitude. The
parameters of these potentials have been adjusted to fit the experimental
data, which consist of the scattering volume, the phase shifts up to the
pion production threshold, and the position of the nucleon pole with the
residue at the pole being the $\pi NN$ coupling constant when all legs
in the vertex are on-mass-shell. The renormalized $\pi NN$ form factor
$f^R(k,E)$, which is energy dependent, is given by~\cite{MA85}
\begin{eqnarray}
f^R(k,E) &=& Z_2^{1/2}(E)\,\big[ f_0(k) \nonumber \\
& & + g_\alpha(k)\,\tau_\alpha(E)\,\langle g_\alpha|\,
G_0(E)\,|f_0\rangle\big] \ ,                                  \label{eq:3}
\end{eqnarray}
with $\alpha=P_{11}$~\cite{Foot1}. Here, $G_0(E)$ is the free $\pi N$
Green's function, $Z_2(E)$ is the wave function renormalization, and $f_0(k)$
is the bare $\pi NN$ form factor. The nonpole part of the $P_{11}$ $\pi N$
amplitude gives, in this case, the dressing to the $\pi NN$ form factor. The
dressed nucleon propagator $d^R(E)$ has a pole with unit residue at $E=m_N$,
i.e., $d^R(E) = (E-m_N)^{-1}$.  The important feature of this $\pi N$ $P_{11}$
amplitude is that the background term
$g_\alpha(k)\,\tau_\alpha(E)\,g_\alpha(k')$ is part of the amplitude that
contributes to $\pi N$ scattering in the three-nucleon force, while the pole
term gives the $\pi NN$ form factor for the pion production and absorption
vertices. In this case the dressed $\pi NN$ form factor $f^R(k,E)$ is not
free, but is constrained by the experimental $\pi N$ data in the $P_{11}$
channel. To that extent the parameters of this three-nucleon force are
completely determined by the $\pi N$ data. The off-shell behavior of this
$\pi N$ amplitude is completely determined by the choice of the bare
$\pi NN$ form factor $f_0(k)$ and the form factors of the separable potential
$g_\alpha(k)$. This completely defines the three-body force corresponding to
the diagram in Fig.~\ref{Fig.1}.

For the present calculation we have taken the triton wave function to be the
solution to the 18 channel Faddeev equation for the Paris-Ernst-Shakin-Thaler
(PEST)
potential \cite{HP84,PK91} which is a separable expansion to the Paris
potential
\cite{LL80}. We have taken sufficient terms in the separable expansion to
get the binding energy, $S$-, $S'$-, and $D$-state probability for the triton
to be 7.318~MeV, 90.111\%, 1.430\%, and 8.393\%, respectively. These are in
very good agreement with the corresponding results from the exact coordinate
space Faddeev calculations \cite{FG88} which give 7.388~MeV, 90.130\%,
1.395\%, and 8.409\%. This establishes that our three-nucleon wave function
is a good approximation to the exact wave function extracted from the
coordinate space Faddeev equations.

In Table~\ref{Table.1}, we present the contribution of the three-body force,
in keV, to the binding energy of the triton from the different $\pi N$
partial waves and for different choices of the interaction in the $P_{11}$
channel. The potentials $PJ$ and $M1$ are those of McLeod and
Afnan~\cite{MA85}. The potential $PJ$ has a dipole bare form factor $f_0(k)
\propto k/(k^2 + \alpha^2)^2$ with $\alpha = 3.82$~fm$^{-1}$ ($\approx
754$~MeV), while potential $M1$ has a monopole bare form factor $f_0(k)\propto
k/(k^2 +\alpha^2)$ with $\alpha=2.77$~fm$^{-1}$ ($\approx 547$~MeV). For
comparison, we have also included results for two potentials $A$ and
$D$\cite{SA94}, which have been adjusted to fit the same $\pi N$ data as the
potentials $PJ$ and $M1$. These have a bare form factor which is a monopole
with a range or cutoff mass of $\alpha = 9.32$~fm$^{-1}$ ($\approx 1840$~MeV)
and $\alpha = 1.64$~fm$^{-1}$ ($\approx 323$~MeV) respectively. From these
results we may conclude that: (i)~The total contribution from all $\pi N$
partial waves is very small, less than 50 keV, and not of any significance in
explaining the discrepancy between the results of Faddeev calculations based
on a two-body $NN$ interaction and the experimental binding energy. (ii)~The
largest contributions are from the $P_{11}$, $P_{33}$, and $S_{31}$ partial
waves. This justifies the fact that the $\Delta(1230)$ plays a large role in
the determination of the three-body force. But then the $S_{31}$ and $P_{11}$
$\pi N$ amplitude are equally important. Here we should point out that the
$S$-wave $\pi N$ interaction used has scattering lengths such that
$a_1+2a_3= -0.011 m_\pi^{-1}$, consistent with the requirement of current
algebra. (iii)~There is a cancellation between the $P$-wave and
$S$-wave contributions, making the overall contribution of the three-body
force even smaller. (iv)~The results are sensitive to the choice of the bare
$\pi NN$ form factor, to the extent that potential $A$ gives a much larger
contribution than potential $D$. However, this variation is not very large
considering the range of cutoff masses. This is a result of the fact that
the corresponding dressed form factors are not drastically different, see
Fig.~\ref{Fig.2}. This is an indication that the requirement to fit the
experimental data puts considerable constraint on the dressed form
factor\cite{Pe94}, and that in turn constrains the variation in the
contribution of the three-body force to the binding energy of the
three-nucleon system. (v)~Finally, there is a correlation between the sign of
the contribution of the different partial waves and the corresponding phase
shift, which is expected considering the fact that the sign of the phase
shift tells us if the interaction in that partial wave is attractive or
repulsive. Here we should note that the $P_{11}$ contribution, which is due
to the nonpole part of the amplitude in this channel, is attractive, as is
the case with the $P_{33}$. This is due to the fact that the nonpole
$P_{11}$ phases are similar in behavior to the $P_{33}$
phases~\cite{MF81,MA82}.

To examine the possible reasons for this small three-body force contribution
to the triton binding energy, and in particular, the role of the energy
dependence of the $\pi N$ amplitude and the cutoff mass in the $\pi NN$ form
factor, we need to consider different approximations to the above results. We
first consider the energy dependence of the dressed $\pi NN$ form factor. We
could take the energy in this form factor to be the nucleon mass, i.e.,
\begin{equation}
f^R(k,E_{\pi N})\rightarrow f^R(k,m_N)\ .                        \label{eq:4}
\end{equation}
The results of this approximation for potential $PJ$, labeled (i) in
Table~\ref{Table.2}, give us some additional binding, but nothing
substantial. In this case both the $S$-wave and $P$-wave contributions have
increased in magnitude, but the cancellation between the $S_{31}$ and the
$P_{33}$ reduces the overall contribution. If in addition we take the energy
in the $\pi N$ amplitude to be fixed at  the nucleon mass, i.e.,
\begin{equation}
\tau_\alpha(E_{\pi N})\rightarrow\tau_\alpha(m_N) \ ,           \label{eq:5}
\end{equation}
then we get a substantial increase in the overall contribution to the
binding, see Table~\ref{Table.2} line labeled (ii). This is partly due to the
fact that the $S_{31}$ contribution has been reduced, while the $P_{33}$ and
$P_{11}$ contributions have increased. As a result, the total contribution to
the binding energy has increased by a factor of ten, when the energy in both
the $\pi N$ scattering amplitude and the dressed $\pi NN$ form factors  are
fixed to be $E_{\pi N}=m_N$. To understand this change in cancellation as we
change the $\pi N$ energy, we should recall that
$\tau_\alpha(E_{\pi N}) = [\lambda^{-1} - \langle
g_\alpha|G_0(E_{\pi N})|g_\alpha\rangle]^{-1}$, where $\lambda$, the strength
of
the
potential, is $-1$ for attractive potentials such as is the case for the
$P_{33}$ channel,  and $+1$ for repulsive potentials as is the case for the
$S_{31}$. Since $\langle g_\alpha|G_0(E_{\pi N})|g_\alpha\rangle$ is negative
for $E_{\pi N}<(m_\pi+m_N)$, then the two terms in the denominator of
$\tau_\alpha(E_{\pi N})$ add for repulsive potentials and subtract for
attractive potentials. Furthermore, since the magnitude of $\langle
g_\alpha|G_0(E_{\pi N})|g_\alpha\rangle$ increases as we approach the $\pi N$
threshold from below, the contribution of the repulsive channels is reduced
while that of the attractive channels increases. In fact if we take the
energy in the $\pi N$ amplitude to be the $\pi N$ threshold, i.e., $E_{\pi
N}=(m_\pi + m_N)$, then the contribution of the three-body force to the
binding energy of the triton increases by another factor of two as
illustrated in Table~\ref{Table.2} line labeled (iii). This establishes the
fact that the inclusion of the energy dependence of the $\pi N$ amplitude and
dressed $\pi NN$ form factor into the calculation has reduced the total
contribution to the binding energy by a factor of $\approx 40$.

We now turn to the role of the $\pi NN$ form factor in determining the
magnitude of the three-nucleon force contribution to the binding energy of
the triton. Here we observe that there is a bare form factor $f_0(k)$ that
can be chosen to be a monopole or a dipole form factor with some fixed
cutoff mass. However, the {\it dressed} $\pi NN$ form factor given in
Eq.~(\ref{eq:3}) and used for the pion absorption and production vertex is
determined by fitting the experimental data. In Fig.~\ref{Fig.2} we compare
the dressed $\pi NN$ form factors for the potentials considered in
Table~\ref{Table.1} with a monopole form factor, i.e., $F_0(k) =
\Lambda^2/(k^2 + \Lambda^2)$, having a cutoff mass of
$400$ and $800$~MeV. Here we observe that the dressed form factors are
almost identical despite the fact that the different potentials have either a
monopole form factor (potentials $M1$, $A$, and $D$) with drastically
different cutoff masses, or a dipole form factor (potential $PJ$). This
strongly suggests that the dressing and the requirement of fitting the
experimental data has the effect of constraining the dressed $\pi NN$ form
factors irrespective of the choice for the cutoff mass in the bare form
factor\cite{Pe94}. Since in previous determinations of the contribution of the
three-nucleon potential to the binding energy of the triton, no experimental
constraint was imposed on the $\pi NN$ form factor other than the
Goldberger-Treiman relation~\cite{CS90} for the TM potential, we now examine
the result of replacing the dressed $\pi NN$ form factors in our model with
the potential $PJ$, by the monopole form factor $F_0(k)$. At the same time we
fix the energy of the $\pi N$ amplitude at the nucleon pole and take
the form factor for the separable potentials, $g_\alpha(k)$, to be also the
monopole form factor $F_0(k)$. This can be achieved by the substitutions
\begin{eqnarray}
f^R(k,E_{\pi N})        &\rightarrow&  \frac{f^R(k,m_N)}{k}\Bigg|_{k=0}
 k\, F_0(k)\ ,  \nonumber  \\
\tau_\alpha(E_{\pi N})  &\rightarrow&   \tau_\alpha(m_N)\ ,
\label{eq:6}\\
g_\alpha(k)             &\rightarrow&    \frac{g_\alpha(k)}{k^\ell}
\Bigg|_{k=0} k^\ell\, F_0(k)\ ,\nonumber
\end{eqnarray}
where $\ell$ is zero for $S$ waves and one for $P$ waves. This procedure is
similar to that implemented in the TM potential~\cite{Foot2}. This
approximation involves the assumption that the form factor in the separable
potentials is the {\it same} in all partial waves, and of the monopole
type\cite{Foot3}. In Table~\ref{Table.2} we present the results of such a
substitution for $\Lambda = 400$ and $800$~MeV. The substitution in
Eq.~(\ref{eq:6}) has the dramatic effect of increasing the contribution of
the three-nucleon force by one to two orders of magnitude, with the final
result being very sensitive to the choice of $\Lambda$. Note that the
difference between the results with $\Lambda = 400$ and $800$~MeV should
be compared with the results for, say, the potentials $M1$ and $A$ in
Table~\ref{Table.1}, which have cutoff masses for the bare form factors of
547 and 1822~MeV. To understand this large change, we recall from
Fig.~\ref{Fig.2} that for $k\approx 3$, the monopole form factor for
$\Lambda=800$~MeV is about three times greater than that of the potential $PJ$,
and this form factor is raised to the fourth power, since we have four
factors of $F_0(k)$, two from the $\pi NN$ form factors, and two from the
separable potential form factors. This should give us an increase by roughly
two orders of magnitude, which is consistent with the result of comparing the
fourth and last line in Table~\ref{Table.2}. On the other hand, by comparing
the results for the potentials $M1$, $A$, and $B$ in Table~\ref{Table.1}, we
observe some sensitivity to the dressed $\pi NN$ form factor. This
sensitivity is not as dramatic partly because the energy dependence of the
$\pi N$ amplitude and the $\pi NN$ form factor result in a cancellation
between the repulsive $S_{31}$ and the attractive $P_{11}$ and
$P_{33}$ contribution, and partly because the dressed form factors for the
different potentials have similar momentum dependence. The role of the
cancellation between the contribution of the different partial waves can be
illustrated by comparing the results for the potential $A$ in
Table~\ref{Table.1}, with those for $\Lambda=400$~MeV in Table~\ref{Table.2},
both of which have comparable $\pi NN$ form factors, see Fig.~\ref{Fig.2}.

Although this analysis is based on perturbation  theory, the fact that the
results are so small suggests that a more exact treatment of the three-body
force based on the present formalism is not warranted since higher order
contributions from the three-nucleon force cannot change the overall magnitude
of the three-body force substantially\cite{Bo86}. Within the framework of
$\pi NN$ dynamics there should be a component of the three-nucleon system
that corresponds to $\pi NNN$. This comes in at higher order if the
perturbation expansion is about three point nucleons. We have chosen not to
include this effect on the grounds that it would most likely give a small
contribution.
The inclusion of the other time order $\pi-\pi$ three-body force can at most
increase the contribution by a factor of $4$, if all contributions add
coherently and are of the same magnitude, which is still too small to bridge
the gap between the experimental result of 8.48~MeV and the value for the
binding energy based on the Paris two-body interaction of 7.39~MeV. In the
$NN-\Delta N$ coupled channel approach there is a cancellation between the
three-body force contribution and the dispersive effect. If this cancellation
persists in channels other than the $P_{33}$, we would expect a further
suppression of the three-body contribution to the binding energy. The present
results are based on the use of separable potential for the nonpole
interaction, and as a result only the $P_{11}$ data put a constraint on the
dressed $\pi NN$ form factor. For chiral Lagrangians, e.g.,~\cite{PJ91}, the
total $\pi N$ data are used to constrain the dressed
$\pi NN$ form factor. We need to establish if this limited variation in the
$\pi NN$ form factor is a special feature of the present model, or is more
generally valid.

In summary, we have found two factors that  suppress the contribution of the
three-nucleon force to the binding energy  of the triton. These are the
following: (i)~The
energy dependence of the $\pi N$ amplitude and the dressed $\pi NN$ form
factor reduce this contribution by $\approx 40$. Part of this effect is due
to the fact that there is a substantial cancellation between the repulsive
$S_{31}$ contribution and the attractive $P_{11}$ and $P_{33}$ contributions.
(ii)~The use of a monopole form factor with a large cutoff mass can enhance
the contribution of the three-nucleon force by as much as two orders of
magnitude provided the energy dependence of the $\pi N$ amplitude and the
$\pi NN$ form factor are neglected.

The authors would like to thank the  Australian Research Council and Flinders
University Board  of Research for their financial support during the course of
this work. The authors would like  to thank J. Haidenbauer for supplying them
with the parameters of the PEST potential, D.R.~Lehman for comparison of
triton results for the PEST potential,  and B.~Blankleider for the code to
generate the potentials $A$ and $D$. Finally we would like to thank
B.F.~Gibson for supplying us with the results of the coordinate space
calculation for the triton with the Paris potential. One of the authors
(I.R.A.)
would like to acknowledge very informative discussions with A.W.~Thomas and
B.C.~Pearce on the subject.

\newpage

\begin{figure}[h]
\caption{The contribution to the three-nucleon force.\label{Fig.1}}
\end{figure}



\begin{figure}[h]
\caption{Comparison of the dressed $\pi NN$ form factors for  the potentials
$PJ$, $M1$, $A$, and $D$ with a monopole form factor having a cutoff mass of
$\Lambda = 400 and 800$~MeV.\label{Fig.2}}
\end{figure}


\begin{table}
\setdec 00.00
\caption{Comparison of the results for the contribution  of the three-body
force to the triton binding energy in keV for two of the $P_{11}$ $\pi N$
potentials of  Ref. [8]. For comparison, we have also included the
results for potential $A$ and $D$ which have monopole bare form
factors with a cutoff mass of $\approx 1822$~MeV and $\approx 323$~MeV,
respectively.}\label{Table.1}
\begin{tabular}{cccccccc}
$P_{11}$ potential &$S_{11}$&$S_{31}$&$P_{11}$&$P_{31}$&$P_{13}$&$P_{33}$&
Total\\ \tableline
 $PJ$ &\dec -4.8 &\dec 26.4 &\dec -8.8 &\dec -3.6 &\dec 4.5   &\dec -16.0
&\dec -2.3 \\
 $M1$ &\dec -5.0 &\dec 28.9 &\dec -15.3 &\dec -2.1 &\dec 6.2  &\dec -22.1
&\dec -9.4 \\
 $A$  &\dec -9.2 &\dec 51.4 &\dec -38.9 &\dec -4.3 &\dec 10.0 &\dec -38.3
&\dec -29.3 \\
 $D$  &\dec -3.8 &\dec 19.3 &\dec -4.9  &\dec -5.2 &\dec 1.4  &\dec -6.7
&\dec 1.0 \\
\end{tabular}
\end{table}
\vskip 0.5 cm

\begin{table}
\caption{ The effect of removing the energy  dependence in the $\pi NN$ form
factor (i), and the $\pi N$ amplitude (ii) and (iii). Also included are the
results of replacing the dressed
$\pi NN$ form factor and separable potential  form factors by a monopole with
a cutoff mass of $400$ and $800$~MeV, while fixing the energy of the
$\pi N$ amplitude to
$m_N$. The total includes the contribution from all $S$- and $P$-wave
$\pi N$ amplitudes. The results in this table are for the $P_{11}$ potential
$PJ$, and all energies are in keV.}\label{Table.2}
\begin{tabular}{c|cccc}
Approx. & $S_{31}$  & $P_{11}$ & $P_{33}$ & Total \\ \tableline
 exact  &\dec 26.4  &\dec -8.8    &\dec -16.0   &\dec -2.3    \\
 (i)    &\dec 31.7  &\dec -13.0   &\dec -22.3   &\dec -6.8    \\
(ii)    &\dec 28.1  &\dec -20.1   &\dec -39.8   &\dec -36.0   \\
(iii)   &\dec 23.4  &\dec -28.7   &\dec -70.3   &\dec -80.9   \\
400     &\dec 18.7  &\dec -58.8   &\dec -311.5  &\dec -231.1  \\
800     &\dec 113.2 &\dec -1146.7 &\dec -3172.5 &\dec -3897.3 \\
\end{tabular}
\end{table}


\begin{references}

\bibitem{IS86} T.~Sasakawa and S.~Ishikawa, Few-Body  Syst., {\bf 1}, 3
(1986), S.~Ishikawa and T.~Sasakawa, {\it ibid.} {\bf 1}, 143 (1986).

\bibitem{FG88} J.~L. ~Friar, B.~F.~Gibson, and G. ~L. ~Payne, Phys. Rev.
C {\bf 37}, 2869 (1988).

\bibitem {CS79} S.~A.~Coon, M.~D.~Scadron, P.~C.~McNamee, B.~R.~Barrett,
D.~W.~E.~Blatt, and B.~H.~J.~McKellar, Nucl. Phys, {\bf A317}, 242 (1979).

\bibitem{CG81} S.~A.~Coon and W.~Gl\"ockle,  Phys. Rev. C {\bf 23}, 1790
(1981).

\bibitem{CP93} S.~A.~Coon and M.~T.~Pe\~{n}a,  Phys. Rev. C {\bf 48}, 2559
(1993).

\bibitem{BG68} G.~E.~Brown, A.~M.~Green, and W.~J.~Gerace,  Nucl. Phys. {\bf
A115}, 435 (1968).

\bibitem{HS83a} Ch.~Hajduk, P.~U.~Sauer, and W.~Strueve,  Nucl. Phys. {\bf
A405}, 581 (1983).

\bibitem{HS83b} Ch.~Hajduk, P.~U.~Sauer, and S.~N.~Yang,  Nucl. Phys. {\bf
A405}, 605 (1983).

\bibitem{Sa86} P.~U.~Sauer, Prog. Part. Nucl. Phys. {\bf 16}, 35 (1986).

\bibitem{PH90} M.~T.~Pe\~{n}a, H.~Henning, and P.~U.~Sauer,  Phys. Rev. C {\bf
42}, 855 (1990).

\bibitem{SS92} A. ~Stadler and P. ~U. ~Sauer, Phys. Rev.  C {\bf 46}, 64
(1992).

\bibitem{PS93} M.~T.~Pe\~na, P.~U.~Sauer, A.~Stadler, and G.~Kortemeyer,
Phys. Rev. C {\bf 48}, 2208 (1993).

\bibitem{PR91} A.~Picklesimer, R. ~A.~Rice, and R.~Brandenburg,
Phys. Rev. C {\bf 44}, 1359 (1991); {\bf 45}, 547 (1992); Phys.
Rev. Lett. {\bf 68}, 1484 (1992); Phys. Rev.  C {\bf 45}, 2045 (1992);
{\bf 46}, 1178 (1992).

\bibitem{CS90} S.~A.~Coon and M.~D.~Scadron, Phys. Rev. C {\bf 42}, 2256
(1990).

\bibitem{AB80} I.~R.~Afnan and B.~Blankleider, Phys. Rev. C {\bf 22}, 1638
(1980).

\bibitem{AM83} Y.~Avishai and T.~Mizutani, Phys. Rev. C {\bf 27}, 312 (1983).

\bibitem{SSF85} P.~U.~Sauer, M.~Sawicki, and S.~Furui, Prog. Theor. Phys.
{\bf 74}, 1290 (1985).

\bibitem{B92} B.~Blankleider, Nucl. Phys. {\bf A543}, 163c (1992).

\bibitem{AM83b} Y.~Avishai and T.~Mizutani, Nucl. Phys. {\bf A393}, 429
(1983).

\bibitem{CCS93} G.~Cattapan, L. Canton and J.~P.~Svenne, Nuovo. Cimento {\bf
106A}, 1229 (1993).

\bibitem{MF81} T.~Mizutani, C.~Fayard, G.~H.~Lamot, and S.~Nahabetian,  Phys.
Rev. C {\bf 24}, 2633 (1981).

\bibitem{MA82} S.~Morioka and I.~R.~Afnan, Phys. Rev. C {\bf 26}, 1148 (1982).

\bibitem{MA85} R.~J.~McLeod and I.~R.~Afnan, Phys. Rev. C {\bf 32},  222
(1985); C {\bf 32}, 1786E (1985).

\bibitem{Th76} A.~W.~Thomas, Nucl. Phys. {\bf A258}, 417 (1976).

\bibitem{AM85} I.~R.~Afnan and R.~J.~McLeod, Phys. Rev.  C {\bf 31}, 1821
(1985).

\bibitem{Foot1} This definition of $Z_2(E)$ is slightly different  from that
of Ref.~\cite{MA85}.

\bibitem{HP84} J. ~Haidenbauer and W. ~Plesas, Phys. Rev. C {\bf 30},  1822
(1984).

\bibitem{PK91} W. ~C. ~Parke, Y. ~Koike, D. ~R. ~Lehman, and L. ~C. ~Maximon,
Few-Body Syst., {\bf 11}, 89 (1991).

\bibitem{LL80} M.~Lacombe, B. ~Loiseau, J. ~M. ~Richard, R. ~Vinh ~Mau, J.
{}~C\^ot\'e, P. ~Pir\`es, and R. ~de ~Tourreil, Phys. Rev. C {\bf 21}, 861
(1980).

\bibitem{SA94} T.Y.Saito and I.R.~Afnan (unpublished).

\bibitem{Pe94} B.~C.~Pearce, 14$^{\rm th}$ International IUPAP Conference on
Few-Body Problems in Physics, Contributed papers, Ed. F. Gross (Collage of
William and
Mary, Williamsburg,1994), p.294.

\bibitem{Foot2} In the TM potential the amplitude is expanded  in powers of
$Q/m_N$, with $Q$ the pion momentum, and terms up to and including quadratic
terms in the pion momentum retained. At the same time a monopole form factor
which is a function of the momentum $Q$ is introduced into the three-body
force.

\bibitem{Bo86} A.~B\"omelburg, Phys. Rev. C {\bf 34}, 14 (1986).

\bibitem{Foot3} This assumes that the form factor for  $\pi N\leftrightarrow
N^*$ is identical to $\pi N\leftrightarrow N$.

\bibitem{PJ91} B.~C.~Pearce and B.~K.~Jennings, Nucl. Phys. {\bf A528}, 655
(1991).

\end{references}
\end{document}